# 3-Aminopropanol: polymolecular structures that determine unique solvation motifs

Yulia V. Novakovskaya

*Chemistry Department of the Lomonosov Moscow State University, Leninskie Gory, Moscow, Russia 119991*

*E-mail: jvn@phys.chem.msu.ru*

**Abstract.** Intermolecular bonding of 3-aminopropanol (3-AP) molecules is discussed in comparison to 2-aminopropanol (2-AP) and 2-aminoethamol (2-AE). The consideration is based on the results of nonempirical quantum chemical simulations of the molecular clusters carried out at the MP2/6-31+G(d,p) level. Particular attention is paid to the formation of variously ordered 3-AP aggregates, which can be doubled or bracelet rings, extended chains, ribbons, or double helices, impossible in the case of any close amino alcohol analogue, but favorable for the solvation of diverse either hydrophilic or hydrophobic species.

## Introduction

It is a common point of view that molecules in an ordinary liquid can by no means be arranged in a more or less ordered regular way. It seems quite natural that thermal motion should rapidly distort or even destroy any order; and any occasional symmetric local polymolecular fragment should be very short living. Such a behavior is considered as characteristic not only of the liquids with non-directional intermolecular bonds of e.g. dispersion nature, but also of those where molecules form directional bonds, the most interesting and practically important of which are hydrogen bonds. The reason for that is also natural, since, being actually directional, hydrogen bonds provide intermolecular binding within relatively broad ranges of their geometrical parameters. The ranges are especially important when the task is to identify positions of protons in a DNA structure. Therefore, these were primarily analyzed in view of this particular task [1, 2], and one of the conclusions was as follows: for D-H...A bonds, the H...A distance should be smaller than 2.5 Å, while the D-H-A angle should be larger than 90.0º. The interaction energy strongly depends on not only the chemical nature of D and A atoms, but also the H...A distance and the deviation of the bond from linearity. However, the above threshold values show that the interaction is still preserved even upon quite a strong distortion of the structure. This peculiarity reflects the flexibility of hydrogen bonds. And this is the actual reason for the aforementioned thermal distortion of H-bond networks, which makes the existence of its relatively ordered segments unlikely. But there is one additional factor that may determine the structure motifs. This is the co-presence of hydrophobic and hydrophilic parts in the interacting molecules; and a sort of a balance between the interactions of both kinds assisted by the steric effects may play a key role and lead sometimes to quite unexpected results. We are going to illustrate this statement by an example of one of the lowest amino alcohol homologues, namely aminopropanol.

3-Aminopropanol (3-AP) is a close analogue of 2-aminoethanol (2-AE) and 2-aminopropanol (2-AP). Like other amino alcohols, all of them can act as efficient solvents due to the presence of both hydroxyl and amino groups. In the two former these are attached to different terminal carbon atoms, while in the two latter, to the neighboring carbon atoms; and two aminopropanol isomers involve the same number of carbon atoms in the skeleton. Undoubtedly, properties of the compounds are determined by the presence of the hydrophilic functional groups, their mutual effect being mediated by the respective number of methylene fragments. Based on these ideas, 3-AP should in some respects be closer to 2-AE, while in other respects, to 2-AP. If one looks at the physical properties of the substances (Table 1), it may seem that all of them are very similar. Then, a question arises whether the solvent effects and the general character of solvation (stabilization of solute particles) should also be similar for all the three amino alcohols.

The problem deserves particular attention because the lowest amino alcohol homologues, namely, aminoethanol and aminopropanol, are widely used nowadays as basic components of mixtures designed for $CO_2$ scavenging [9–12] or impurity extraction from ores [13]. The solvability of any compound or mixture strongly depends on the individual characteristics of the compounds. Typically, a higher and stronger solubility is reached when solute particles can be tightly surrounded with the solvent ones, and the latter can be spatially ordered due to some peculiar intermolecular interactions, which hampers their mutual mobility and, hence, stabilizes the overall local structure.

**Table 1.** Physicochemical properties of 2-AE, 2-AP, and 3-AP: melting ($T_m$) and boiling ($T_b$) points [3], supercooling temperature range ($\Delta T_{sc}$) [4], heat of evaporation ($\Delta_{vap}H$) [5], density ($\rho_{298}$), viscosity ($\eta_{298}$) and compressibility ($\beta_{298}$) [4]

|  | 2-AE | 3-AP | 2-AP |
|---|---|---|---|
| $T_m$, °C | 10.3 | 11 | 8 |
| $T_b$, °C | 170 | 187 | 174.5 |
| $\Delta T_{sc}$, °C | 38 | 43 | |
| $\Delta_{vap}H$, kJ mol$^{-1}$ | 49.83 | 49.59 | |
| $\rho_{298}$, kg l$^{-1}$ | 1.012 [3, 6]; 1.016 [8] | 0.988 [7,8] | 0.943 [3] |
| $\eta_{298}$, mPa·s· | 18.91 [6]; 26.11 [8] | 40.42 [7]; 44.44 [8] | |
| $\beta_{298}·10^{11}$, Pa$^{-1}$ | 39.6 | 38.9 | |

It is another common point of view that even instantaneous structures that may locally appear in a liquid can be judged only on the basis of some dynamic simulation. The idea seems natural, since the overall thermal excitation of a large molecular ensemble affects any its local fragment. However, here we face a difficultly solvable or even unsolvable problem. To carry out a dynamic simulation of a large molecular ensemble, one should use a semiempirical potential that provides a more or less correct description of the interfragment interactions within and between the molecules. Such potential can be parameterized only on the basis of some quantum chemical simulations; but still always suffers of averaging effects. For example, the selected atomic charges exclude the possibility of molecular self-tuning according to the peculiarities of the electron density redistribution. Furthermore, the mechanically additive schemes in all the potentials can by no means reproduce collective effects related to the local stabilization of homodromic H-bonded rings [14]. Finally, the local fragments of large molecular ensembles retrieved from dynamic trajectory runs are usually close to the corresponding parts of medium-sized or even small molecular clusters where the fragment of interest is surrounded with at least one shell of molecules. Thus, basic features of the local organization of molecules, especially their possible ordering, in a liquid phase can be judged based on the data about the structure organization of molecular clusters.

**Simulation methods**

To suggest a theoretical comparison of the relative solvability of aminoethanol and aminopropanol, we have undertaken a series of nonempirical simulations aimed at distinguishing local structures of their molecular ensembles. Simulations were carried out at the MP2 level (second order of the Moeller-Plesset perturbation theory), which has proved to be a reasonable quantum chemical tool for modeling H-bonded systems where hydroxyl and amino-groups play the key role. Moreover, it enables one to reproduce quite reliably the dispersion-kind interactions between methylene groups. Gaussian-type 6-31G basis set extended with polarization functions on all nuclei is sufficiently flexible for describing diverse isomers of the molecules and relatively compact to avoid the linear dependence. For example, results obtained for 3-AP and 2-AE monomers at the gold-standard CCSD(T)-F12a/cc-pVDZ-F12 level [15] were later shown to be well reproduced at the MP2/6-311G++(d,p) level, the latter structural parameters being in good agreement with the high-resolution rotational spectra of different isotopologues of 3-AP [16]. Taking into account that in molecular clusters, there are numerous functions of the neighboring atoms that provide the necessary flexibility accessible in the case of monomers only in the presence of additional diffuse functions, these extra functions can be excluded from the atomic basis intended to be used for describing cluster systems, especially taking into account the partial linear dependence which rapidly becomes more pronounced with an increase in the cluster size.

The correspondence of the discovered structures to the minima of the potential energy surfaces was confirmed by the normal-coordinate analysis. Vertical ($D_{e,vert}$) and adiabatic ($D_0$) dissociation energies of the clusters (composed of $k$ monomer units) were estimated with the use of 50% counterpoise correction for the basis set superposition error (*BSSE*) and (in the case of $D_0$) with the zero-point energies (*ZPE*) taken into account:

$$D_{e,vert} = \sum_i^k E'_{mon,i} - E_{cluster} - 0.5 BSSE$$

$$D_0 = \sum_i^k (E_{mon,i} + ZPE_{mon,i}) - E_{cluster} + ZPE_{cluster} - 0.5 BSSE$$

$$BSSE = \sum_i^k (E'_{mon,i} - E'^{full}_{mon,i})$$

where *mon,i* and *cluster* subscripts denote *i*th monomer and the whole cluster respectively; *E* and *E'* values stand for the energies of the structure element at its individual optimum configuration and its configuration within the cluster considered; and *full* superscript reflects that the energy is estimated with the use of the basis set of the whole cluster. Thermal contributions to the energy ($G_{rel}$ relative Gibbs energies of the systems of the same atomic composition) were estimated in terms of the statistical thermodynamic approach taking into account vibrational partition functions solely, which tentatively reflects the restricted mobility of particles in a liquid. To judge the spatial organization of the clusters, their apparent surface area (*S*) and volume (*V*) were estimated as the external surface and volume of the overlapping van-der-Waals spheres of all the atoms involved in the system (with the radii of 1.2, 1.5, 1.7, and 1.6 Å for H, O, C and N atoms respectively). The latter aspect is rarely addressed, but seems very important when one aims at identifying possible cluster-like building blocks of a dynamic liquid structure, because the relative energy values (or dissociation energies) of clusters reflect their stability, whereas their relative spatial (geometrical) characteristics indicate the possibility of their particular agglomeration in a relatively dense phase.

Simulations were carried out with the use of Firefly 8.2 program package [17] and visualized with Chemcraft software [18].

**Results and discussion**

In all AP and AE molecules considered, the hydrophilic hydroxyl and amino groups can form a hydrogen bond that closes the molecular structure into a ring. In 2-AE and 2-AP, the groups are attached to the neighboring carbon atoms; and the resulting structural ring is formally five-membered (N, H, O, and two C atoms). In 3-AP where the groups are separated with three methylene fragments, the resulting ring is formally six-membered. The strain of the hydrogen bond formed and, hence, of the whole structure naturally decreases with an increase in the ring size irrespectively of whether the bond is of N-H...O or O-H...N kind. In the ring isomers, the corresponding bond angle increases from 114.5 to 115.2 and 140.6° on going from 2-AP to 2-AE and 3-AP in the case of O-H...N bond (Fig. 1a) and from 100.1 to 102.1 and 128.3° in the case of N-H...O bond (Fig. 1b), respectively. Thus, in agreement with [15, 16], the preferable intramolecular H-bond is of O-H...N kind; it is least strained in 3-AP, while the replacement of one H atom in 2-AE with a $CH_3$ group (which produces 2-AP) causes a weak distortion of the system. In the latter case, the group looks like the last wheel of a coach that tends to reside at a maximum distance from the mean CCON plane of the residual 2nd row atoms in both isomers (Fig. 1).

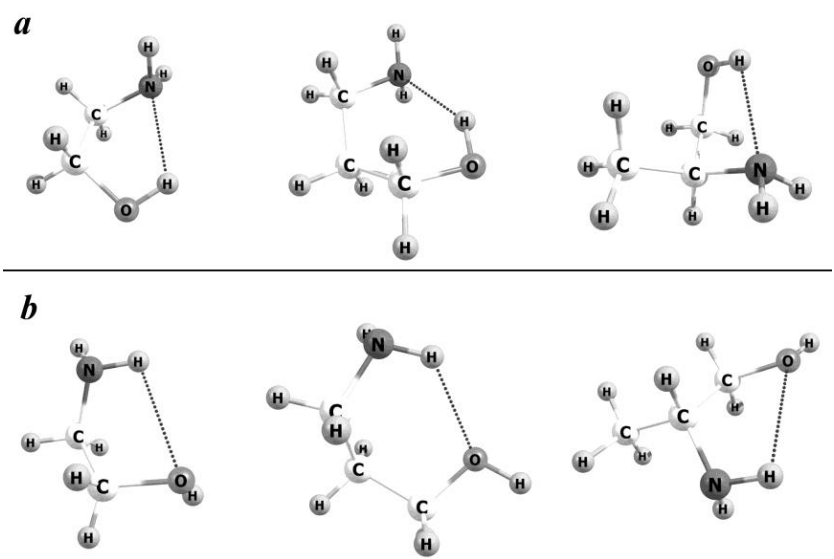

**Figure 1.** Ring isomers of 2-AE, 3-AP, and 2-AP molecules stabilized by intramolecular (a) O-H...N and (b) N-H...O hydrogen bonds.

Energy estimates expectedly confirm the observations, namely, the stabilization energies of the ring isomers that involve O-H...N and N-H...O bonds with respect to the extended isomers where hydroxyl and amino groups are the most distant from each other equal (at 0K) 3.3 and 1.2 kcal/mol in the case of 2-AP; 3.9 and 1.4 kcal/mol in the case 2-AE; and 5.2 and 1.2 kcal/mol in the case of 3-AP, which shows that N-H…O bonds are of nearly the same weak character in all the systems, while O-H…N bond becomes noticeably stronger with an increase in the ring size. Thermal correction, which does not alter the relative stability of isomers, nevertheless causes certain changes, namely the relative Gibbs formation energies of the isomers at 298 K are as follows: -2.3 and -1.0 kcal/mol for 2-AP; -3.0 and -1.1 kcal/mol for 2-AE; and -4.0 and -1.0 kcal/mol for 3-AP. Hence, the key factor that determines the arrangement of amino alcohol molecules is the relative number of mediating methylene groups between the hydrophilic ones and the presence of additional side hydrophobic chains, which weaken the $NH_2$...OH interaction to a certain degree. A sort of balance between the flexibility of a hydrocarbon skeleton and the strength of the intramolecular interaction is reached in 3-AP, the closed molecular rings of which can survive in large molecular ensembles and determine the peculiar coordinating ability. The bond energies of the next homologue, namely, 4-aminobutanol (4-AB), further support the idea. The stabilization and Gibbs formation energies of its closed isomer with an O-H...N bond are only slightly larger than those of the 3-AP isomer, about 5.8 and -4.2 kcal/mol respectively. At the same time, the 4-AB isomer with an N-H...O bond is already unstable at 298K even in a phase with a restricted mobility of molecules. This means that the overall positive effect of the higher flexibility of the longer hydrophobic chain between the hydroxyl and amino groups is nearly maximum in the case of 3-AP: the longer polymethylene bridge makes the mutual coordination of the hydrophilic terminal groups less efficient.

The formation of molecular aggregates is naturally driven by the intermolecular joining of hydrophilic groups via hydrogen bonds at a concurrent certain structuring of the hydrophobic molecular segments. The joining should energetically be most favorable when sequences of alternating covalent and hydrogen bonds of a general Y-H...X-H...Z-H... kind (where X, Y, and Z are electronegative O and N atoms) appear, which (according to our previous studies [14]) provides the conjugation of hydrogen bonds and the overall stabilization of the system.

As was found [19], when two 2-AE molecules approach each other to form a cyclic dimer, their intramolecular bonds are broken, and two strong intermolecular O-H...N joints appear instead (Fig. 2a), which corresponds to the general stabilization effect of ~10 kcal/mol. When intramolecular O-H...N hydrogen bonds are retained in both 2-AE molecules and two intermolecular N-H...O bonds appear (with r(H...O)~2.3 Å), the dimer energy is higher by 2 kcal/mol. In an open structure where both monomers are stabilized by intramolecular O-H...N bonds and one intermolecular N-H...O bond is formed so that a conjugated O-H...N(H)-H...O-H...$NH_2$ chain appears, the total stabilization energy is lower by another 1.5 kcal/mol. This peculiarity is reflected in the structure motifs of large $(OH(CH_2)_2NH_2)_n$ clusters (n = 4–20), the most typical of which are conjugate chains of covalent and hydrogen bonds of many 2-AE molecules and rings that involve two or three molecules [20]. In the whole structure, one cannot distinguish closed elements, because it is branched chains with triangular joint segments formed by two or three 2-AE molecules that prevail.

It is worth noting that the aforementioned features are determined chiefly by the equal number of hydrophilic and hydrophobic groups in 2-AE. As a result, a relatively loose (judging from the number of H-bonds per unit volume) H-bond network with extended continuous (conjugated) segments is formed. Hydrophobic $(CH_2)_2$ segments are built in the network. A certain flexibility of the system is provided by the coexistence of gauche and trans (anti) conformers of 2-AE molecules.

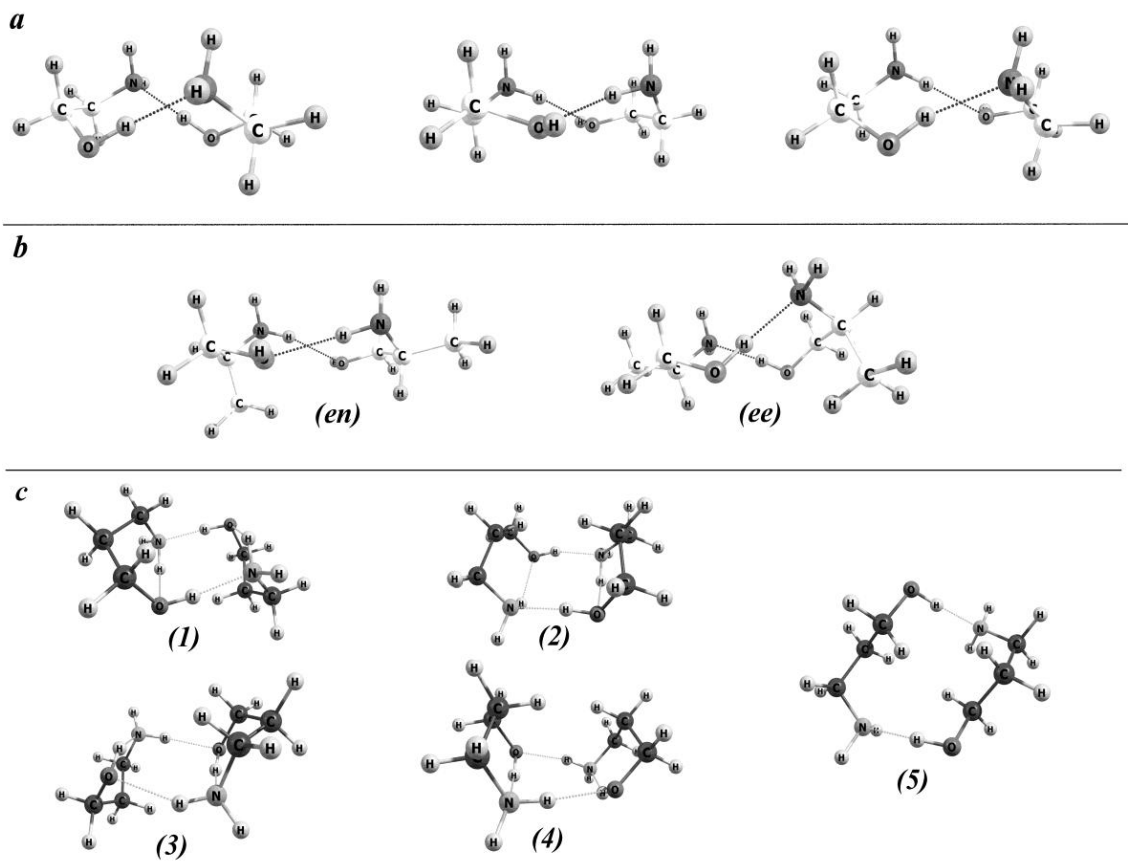

**Figure 2.** Energy-minimum structures of (a) 2-AE, (b) 2-AP, and (c) 3-AP dimers.

The appearance of the third hydrophobic methylene group gives rise to a sort of unbalance between the effective total volumes of hydrophilic and hydrophobic groups, which makes their "uniform" distribution in space less probable, while local aggregation of groups of the same nature more likely. When the group is the aforementioned fifth wheel of a coach as is the case of 2-AP, the main trends typical of 2-AE are preserved at a certain distortion of the whole network because of the presence of these additional $CH_3$ groups. In fact, structure features of 2-AP molecular dimers (Fig. 2b) are similar to those of 2-AE, namely, O-H...N intramolecular bonds typical of the most stable 2-AP monomer do not survive in the most stable dimer naturally stabilized by the intermolecular O-H...N contacts. The resulting eight-membered ring is twisted like in the case of 2-AE, and extra methyl groups reside either in the conditionally equatorial (mean ring) plane of the whole structure (*e*) or normally to it at a side opposite to the NH bonds (*n*). These variations correspond to a relatively narrow energy range, namely, the $D_{e,vert}$ values equal 14.6, 13.9, and 13.3 kcal/mol for the *nn*, *en*, and *ee* configurations of the dimer. The relative Gibbs energies of the three structures also differ by only 0.9 kcal/mol though at a slightly changed order: $G_{rel}(nn) < G_{rel}(ee) < G_{rel}(en)$, being nearly equally spaced in the energy range. Thus, the fifth wheel can actually play a certain stabilizing role when it does not spatially hamper the formation of a hydrogen-bond network, the building blocks of which are the same as in 2-AE.

      The situation changes drastically when one turns to the 3-AP dimer. Here, the formation of intermolecular hydrogen bonds does not cause the breakage of the intramolecular ones. Furthermore, because of the longer hydrocarbon chain between the hydrophilic groups, the latter can be arranged so that a nearly planar NONO quadrangle appears, while the methylene fragments act as a stabilizing support (Fig. 2c). The support can be of a block or step kind. Already this feature shows that there is a real diversity in the 3-AP dimer configurations compared to the 2-AP and 2-AE dimers. The total electronic energies of the structures shown in Fig. 2c fall in a narrow range of 1.7 kcal/mol, the lowest energy being typical of the system where the hydrocarbon step supports the central hydrophilic part with O-H...N intermolecular and N-H...O intramolecular bonds. The stability of the configurations with respect to the dissociation onto the constituting monomers naturally differs because the monomers themselves have different energies and the bonds broken are of different kinds, either stronger O-H...N or weaker N-H...O ones. Nevertheless, as follows from Table 2, both vertical and adiabatic dissociation energies are close for the dimers stabilized by the same kinds of intermolecular bonds but differing in the support kind. The O-H...N bonds provide nearly twice as strong bonding compared to N-H...O contacts. An intermediate dissociation energy is attained in a dimer composed of two monomers with no intramolecular bonds but bonded via O-H...N intermolecular contacts; however in the absence of a conjugated H-bond sequence, the configuration should be distorted by thermal motions most noticeably (judging from the $G_{rel}$ values).

**Table 2.** The dissociation ($D_{e,vert}$ and $D_0$) and relative Gibbs ($G_{rel}$) energies (kcal/mol) of 3-AP dimers shown in Fig. 2c

| Conf. No. | $D_{e,vert}$ | $D_0$ | $G_{rel}$ |
|---|---|---|---|
| 1 | 21.2 | 15.5 | 159.8 |
| 2 | 17.7 | 15.2 | 160.3 |
| 3 | 11.6 | 8.5 | 160.2 |
| 4 | 11.2 | 9.4 | 159.5 |
| 5 | 15.7 | 11.9 | 164.9 |

Note that the strongest O-H...N hydrogen bonds provide an overall similar stability of the chair (step support) and boat (block support) dimer configurations, which means that there is quite a sufficient flexibility (or variability) in the mutual orientation of relatively solid building blocks (which are closed monomer rings). Additionally, a connecting H-bonded N(H)O(H)N(H)O(H) quadrangle of alternating N-H...O and O-H...N bonds, which form a closed conjugated sequence, provides a nearly optimum character of the electron density distribution and, hence, the stability of dimeric units. Then, the question is how the units can be arranged in larger systems. This is well illustrated by the 3-AP hexamer configurations.

A hexamer can formally be composed of three dimers. If all the dimers are boat-like, the largest number of bonds between them is formed when they are arranged in a cycle with the resulting nearly $C_3$ symmetry; and the overall stability is the highest when the H-bonds form a conjugated sequence, in which electronegative O and N atoms are alternating. Such structure (***a***) is shown in Fig. 3. In fact, it is very stable and quite compact (Table 3). All intramolecular O-H...N bonds have lengths of an average of 1.75 Å; intermolecular N-H...O bonds within the rings (i.e. between the dimeric units) are 2.05 Å long and those between the rings (i.e. within the dimers) are 2.13 Å. Thus, the inter-dimeric contacts are even stronger compared to those within the dimers themselves. Therefore, under natural conditions, molecules by no means should assemble by keeping their favorable pair-wise coordination to each other. There should be cooperative effects; and one of such is the consistent shift or rotation of some molecules that may result in the general stabilization of the system. In fact, the arrangement of molecules in structure ***a*** does not provide broad possibilities for further growth of the system in a complementary way. The increase may take place in an easier way when the molecular rings are rotated with respect to each other to give the resulting symmetry close to $D_3$.

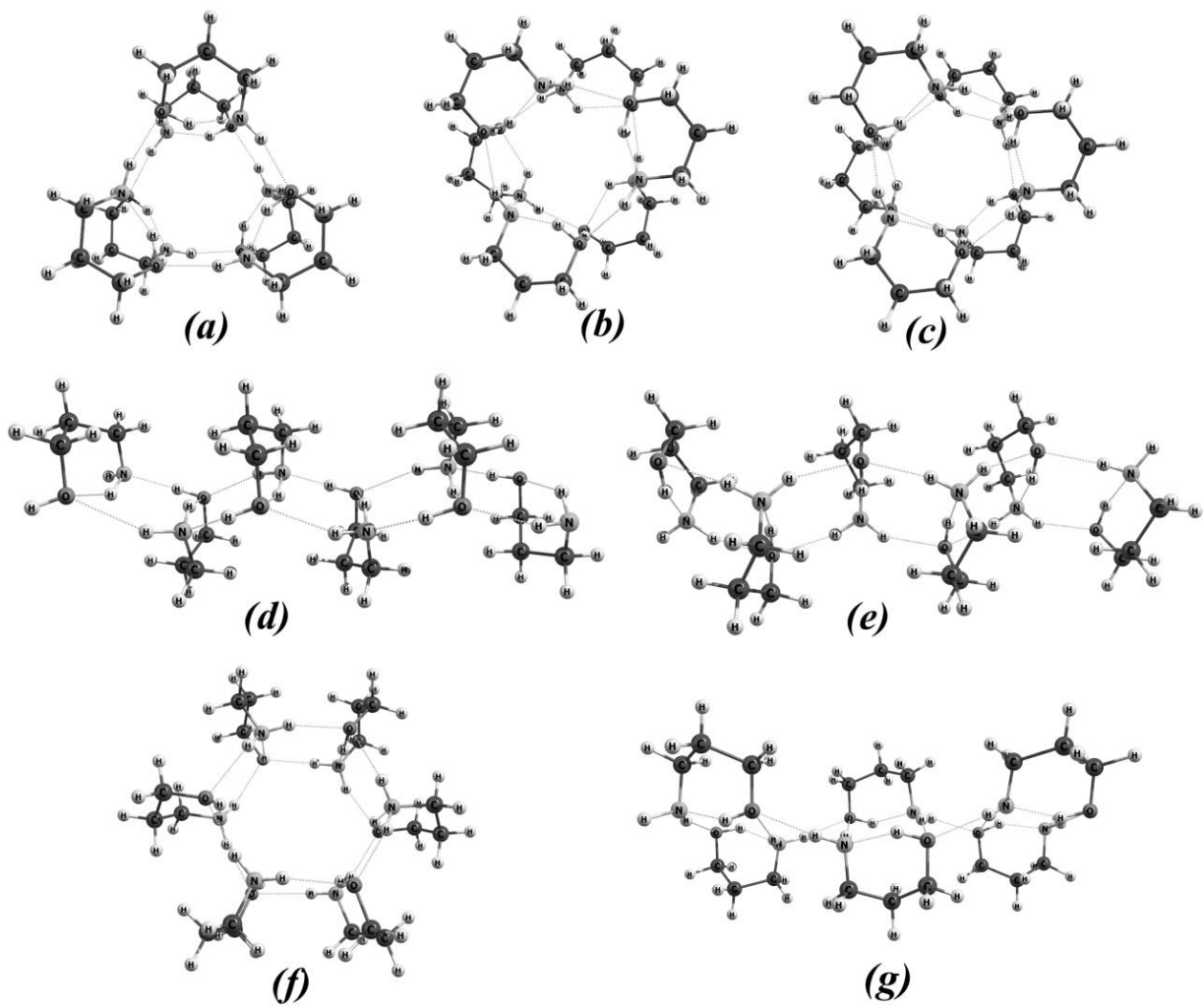

**Figure 3.** Most stable regular structures of 3-AP hexamer.

**Table 3.** The dissociation ($D_{e,vert}$) and relative Gibbs ($G_{rel}$) energies (kcal/mol), apparent surface areas ($S$, Å$^2$) and volumes ($V$, Å$^3$) of the regular 3-AP hexamer structures shown in Fig. 3

| Conf. No. | $D_{e,vert}$ | $G_{rel}$ | $S$ | $V$ |
|---|---|---|---|---|
| **a** | 61.9 | 463.0 | 616.9 | 477.3 |
| **b** | 67.6 | 465.0 | 620.1 | 478.3 |
| **c** | 63.0 | 461.6 | 615.8 | 476.8 |
| **d** | 61.4 | 485.1 | 640.0 | 481.5 |
| **e** | 49.6 | 472.4 | 631.0 | 479.9 |
| **f** | 52.4 | 471.2 | 623.1 | 479.9 |
| **g** | 49.4 | 470.6 | 632.1 | 479.0 |

One such structure is shown in Fig. 3 (***b***). Because of the formal rotation of the lower ring by 60º with respect to the upper one, the alternation of O and N vertices noticed in structure ***a*** between the rings is lost (while kept in the rings themselves), so that intermolecular contacts between the rings become either stronger O-H...O (1.82 Å) or weaker N-H...N (2.01 Å) ones, which makes the whole structure slightly wavy. Concurrently and inevitably the intramolecular bonds in the lower ring become also of N-H...N (2.21 Å) kind similarly to the intermolecular ones (2.19 Å), which results in a general unification, but weakening of all H-bonds in one trimolecular ring because of the relatively strained local coordination neighborhoods of amino groups. In complete agreement with the above discussion, such reorganization results in an increase in the dissociation energy, but at the same time in an increase in the relative Gibbs energy as well, which means that the structure is relatively tightly bonded but can be distorted thermally.

A general improvement in the local organization is reached when all intramolecular bonds are of O-H…N kind at the concurrent formation of intermolecular N-H…O bonds so that the alternation of O and N atoms within a conjugated H-bonded sequence, which spans the whole cluster, is also restored (Fig. 3, configuration *c*). This cluster configuration can by no means be obtained by combining dimeric units, but it preserves all the key features distinguished in them. The intramolecular O-H...N bonds are again the shortest ones (1.75 Å on the average); all intermolecular bonds are of N-H...O kind, being a little bit shorter within the rings (2.07 Å) than between them (2.09 Å). On the average, the H-bond contacts between the rings are larger in this isomer compared to *b* structure; but the intermolecular bonds in cluster *c* are more structurally and energetically uniform. As a result, the vertical dissociation energy of *c* structure is slightly lower than that of *b*, but the thermal distortion of the latter should be less pronounced (Table 3), which is very important if one considers the systems as precursors of sufficiently long-living liquid-phase structural fragments.

It is high time to recall that along with the boat-like dimers, chair-like ones were also identified. At first sight, it seems difficult to arrange them in a closed structure, but they can easily be joined in a chain as shown in Fig. 3 (structure *d*). When a closed ring is opened and transformed into a chain, the possibilities for the general thermal distortion become broader, which is naturally reflected in the noticeably higher relative Gibbs energy (Table 3), a mean molecular contribution to which becomes higher by almost 4 kcal/mol. However, the dissociation energy is at the same high level as for the $C_3$-like ring structure despite the smaller number of hydrogen bonds. The *d* chain is stabilized by the equal number of intermolecular O-H…N and N-H...O bonds, while all intramolecular contacts are of N-H...O kind. A good portion of the thermal energy increment is due to this very peculiarity, namely (i) the stronger O-H...N bonds (1.82 Å in the inner part of the chain) are so-to-say partly counterbalanced by the weaker N-H...O ones (2.27 Å), since they are always complementary and opposing each other within each chain link, and (ii) the weaker intramolecular bonds themselves (2.13 Å) make each link of the chain less rigid.

If all the molecules in such a ribbon are stabilized by O-H...N intramolecular bonds, as in hexamer structure *e* (Fig. 3), their internal configuration is more rigid, while their bonding to each other is naturally weaker as provided by N-H...O bonds. Lengths of all the latter fall in a range of 2.09 to 2.17 Å; and all O and N atoms except the terminal ones are symmetric nodes, O atoms beings double H-acceptors, while N atoms acting as double donors to the left and right neighbors of the ribbon units. Such coordination makes the N-H...O bonds relatively strained, which is inevitably reflected in the lower dissociation energy of this isomer compared to *d*, namely, the $D_{e, vert}$ value is smaller by a dozen kcal/mol. At the same time, the thermal energy increment judged from the $G_{rel}$ value is smaller in the case of *e* isomer, because the internal structure of the ribbon units (3-AP molecules) stabilized by stronger O-H...N bonds is more rigid and less distortable. In fact, the relative Gibbs energy of the isomer is intermediate between the aforementioned rings and a ribbon chain *d*. The same can be said about the surface area of the structure, namely, it is larger than those of rings, but smaller than that of *d* chain (Table 3).

In fact, boat-like dimers can be organized in a similar way, but (by contrast to the above structure which is spatially extended) their joining should inevitably lead to the formation of a noticeably curved chain, which can be closed already at the number of molecules equal to six, as shown in Fig. 3 (configuration *f*). This cycle substantially differs from those discussed earlier; though its general symmetry is still close to $C_3$. Now, the molecular skeletons act as outer walls of the whole cycle (or cyclic cylinder); O-H...N intramolecular bonds (1.77 Å on the average) form the inner walls; while the upper and lower rings are all composed of N-H...O bonds, whose lengths are consistently alternating in both upper and lower rings, being 2.14 and 2.25 Å within the neighboring segments. As a result, any H-bonded O...N...O...N side quadrangle is characterized by a conjugated sequence of bonds, which adds to the overall stability of the cluster despite the relatively strained intermolecular N-H…O bonds. This peculiarity predetermines the dissociation energy close to that of isomer *e* (by almost 10 kcal/mol lower compared to isomer *d*). And, which is even more important, the thermal distortion of the cycle is smaller than that of the open chain *d*, but nearly equal to that of chain *e*; and the cycle itself is reasonably compact (Table 3).

Another modification, which results in a slight additional decrease in the dissociation energy at a comparable concurrent decrease in the relative Gibbs energy, is the side connection of chair-like dimers as shown in Fig. 3 (structure *g*). In this structure, the H-bonded sequence looks like a segment of a double helix composed of quadrangular segments, in which the role of sewing intra- and interhelical joints belongs to intermolecular N-H…O bonds (of 2.10 and 2.20 Å on the average respectively), while the main stabilizing role is played by the intramolecular O-H…N bonds (1.81 Å on the average) and the respective electron density redistribution provided by the alternation of O and N electronegative atoms in the H-bond sequences. As a result, the structure has an effective volume comparable to those of the most compact rings and a larger surface, which corresponds to a larger contact area with the neighbors in a condensed phase. At the same time, the thermal energy increment of the structure is noticeably lower than that of chain isomer *d*, but higher (though not as strongly) than those of all rings.

Undoubtedly, additionally to all these regular and more or less symmetric hexamer structures, there are vast isomers with less regular configurations, which involve hydrophobic contacts as well (Fig. 4). If there are two such contacts as in isomer *h*, which comprises a dimeric chair-like unit and a tetrameric unit composed of two boat-like dimers, one can notice a typical arrangement of methylene groups of the neighboring molecules spatially shifted with respect to each other. Because of the smaller number of hydrogen bonds (12 compared to 16 or 18 in the above isomers), the dissociation energy is naturally lower even than that of ribbon-like isomers, but the effective volume and surface area of the cluster are close to those of isomer *e* (Table 4). The relative Gibbs energies of these two isomers are also close. On the whole, this means that segments of compact ring-like isomers can form hydrophobic contacts between each other, and the resulting configurations are also quite stable at a level typical of extended ribbon-like structures.

When the number of intermolecular hydrogen bonds is further decreased, as in isomers *i* and *j* (Fig. 4), all the bonds are of N-H...O kind, and there are already two so-to-say three-vertex hydrophobic contacts, the overall stability of the cluster expectedly decreases at a concurrent increase in the effective volume and the surface area. Here, when 3-AP monomers are stabilized with weaker N-H...O bonds, while intermolecular bonds in boat-like dimers are quite strong O-H...N ones, the dissociation energy of the cluster is higher (isomer *i*) compared to the configuration with the inverse character of intra- and intermolecular hydrogen bonds (isomer *j*). However, it is important to stress that the relative Gibbs energies of the clusters are close and inverted compared to the $D_{e,vert}$ values (Table 4). Thus, one can see again that clusters composed of 3-AP molecules with O-H...N intramolecular bonds are preferable from the point of view of both their spatial organization (compactness) and thermal stability.

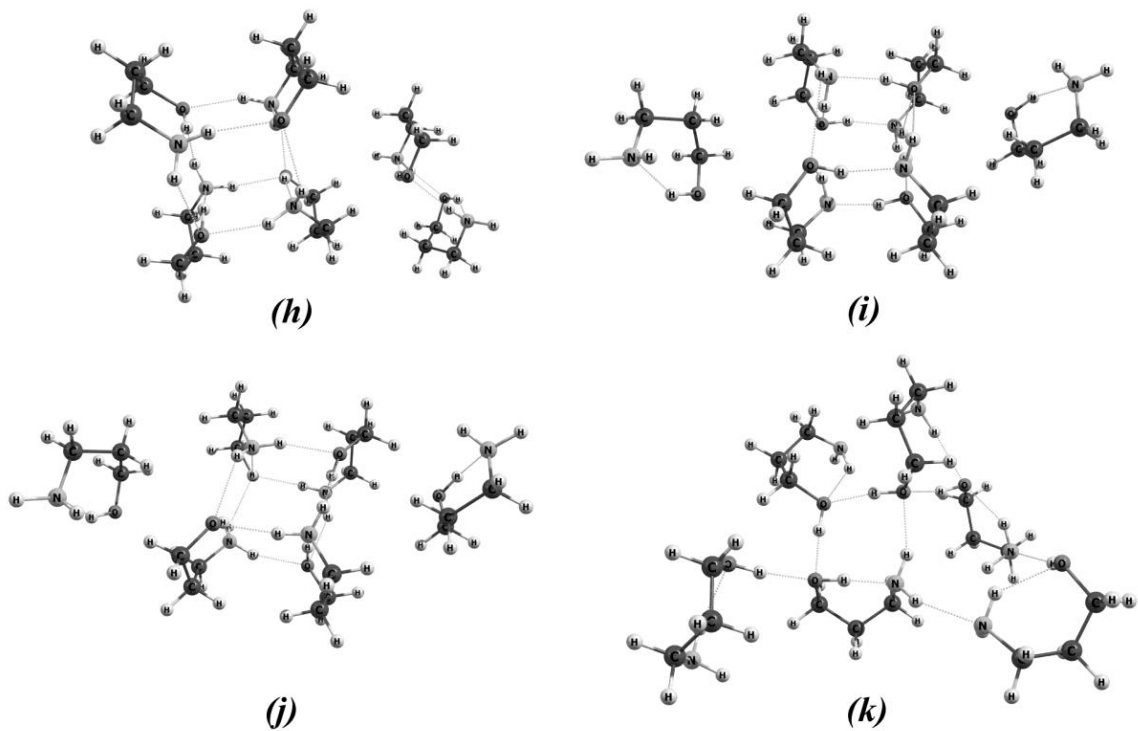

*(h)* *(i)* *(j)* *(k)*

**Figure 4.** Irregular structures of 3-AP hexamer.

**Table 4.** The dissociation ($D_{e,vert}$) and relative Gibbs ($G_{rel}$) energies (kcal/mol), apparent surface areas ($S$, Å$^2$) and volumes ($V$, Å$^3$) of the irregular 3-AP hexamer structures shown in Fig. 4

| Conf. No. | $D_{e,vert}$ | $G_{rel}$ | $S$ | $V$ |
|---|---|---|---|---|
| **h** | 45.6 | 473.3 | 635.0 | 479.8 |
| **i** | 49.0 | 482.4 | 641.5 | 483.4 |
| **j** | 35.6 | 480.5 | 644.5 | 482.3 |
| **k** | 56.3 | 486.0 | 641.3 | 483.4 |

Finally, let us consider an example with no regularity in the mutual arrangement of molecules. Numerous structures of this kind can and were actually generated starting from arbitrary spatial positions and orientations of 3-AP molecules. Figure 4 shows one such structure referred to as ***k***. One can notice here molecules with either N-H...O or O-H...N intramolecular bonds; while intermolecular contacts are of N-H...O (2), O-H...N (1), O-H...O (4), and N-H...N (1) kinds. As in 2-AE clusters [19, 20], one can distinguish here conjugated sequences of H-bonds that collectively span the whole cluster, but now involve both inter- and intramolecular bonds. The conjugation effect contributes to the overall stability of the cluster at a low temperature ($D_{e,vert}$ value, Table 4), but the relatively small number of hydrogen bonds (13), among which there are very weak ones, makes the structure easily distortable ($G_{rel}$ value). It is worth noting that the spatial characteristics of the cluster fall in the range typical of all irregular isomers shown in Fig. 4.

This means that when one carries out a molecular dynamics simulation with an empirical or semiempirical potential, which is parameterized to reproduce some bulk characteristics of the compound studied, such as its density, there is no criterion to check whether the mutual arrangement of molecules crucial for any solvation process is predicted correctly or there is just a compensation of opposite effects that provide the resulting averaged parameter close to the target one. In our opinion, the results considered in this paper shed light on the peculiarities of intermolecular interactions in one particular case, namely in ensembles of 3-aminopropanol molecules, which is very essential in view of the involvement of 3-AP in solvation of diverse compounds either as an individual substance or as a mixture component.

## Conclusions

Thus, what do we see in relatively small 3-AP clusters and what can be expected in a condensed phase? All the hexamolecular structures considered in detail in this work are characterized by the comparable high stability against dissociation onto constituting molecules. The stability is predetermined by the large number of conjugated hydrogen bonds. At the same time, most of the intermolecular bonds in the most stable aggregates are of N-H…O kind and can be thermally distorted to an extent sufficient for the local structure reorganization. Thus, not a high external perturbation (produced by the appearance of a foreign particle) can provide the necessary local reorganization that favors the solvation of this particular solute. Simultaneously, the diversity of structures illustrates the broad range of mutual arrangements of molecules. The symmetry of the arrangements in individual 3-AP ensembles can further be stabilized by foreign particles. In other words, depending on the configuration (or electrostatic potential) of the foreign particle, it can initiate the appropriate rearrangement of 3-AP molecules. The resulting structure motifs may be either extended chains, or double helices, or closed rings with different electrostatic-potential steps complementary with the particular foreign particle. Individual double-ring structures (like hexamer isomers *a*, *b*, and *c*) can solvate small hydrophilic particles, while stacks of such doubled rings can solvate hydrophilic chain-like species. Bracelet rings (like isomer *f*) formed by a sequence of individual molecules whose hydrophobic skeletons are parallel to the ring axis can solvate larger hydrophilic species. All the rings are characterized by the smallest surface areas and volumes (among the isomers analyzed), which reflects their compactness and, at the same time, the possibility of a certain swelling if necessary for the proper localization of a solute particle. Extended chains (like isomers *d* and *e*) can serve as mediating interlayers between hydrophobic species; while double-helical structures (like isomer *g*) can act as efficient interlayers between DNA molecules or any other polypeptide chains. The two latter isomers are characterized by the largest (among the structures considered) surface areas which provides the best possibilities for the proper extended contact with a solute. In this respect, 3-aminopropanol is a unique molecule; and none of the close analogues can provide anything of this kind or even comparable.

## Acknowledgments

The work was supported by the Russian Foundation for Basic Research, project no. 19-03-00215.

## References

[1] I.K. McDonald, J.M. Thornton, Satisfying Hydrogen Bonding Potential in Proteins, *J. Mol. Biol.* 238 (1994) 777-793. https://doi.org/10.1006/jmbi.1994.1334.

[2] I.Y. Torshin, I.T. Weber, R.W. Harrison, Geometric criteria of hydrogen bonds in proteins and identification of `bifurcated' hydrogen bonds, *Prot. Eng. Design and Selection* 15 (2002) 359-363. https://doi.org/10.1093/protein/15.5.359.

[3] Acros Organics. Catalogue of Fine Chemicals. 2002-2003.

[4] I.A. Solonina, M.N. Rodnikova, M.P. Kiselev, A.V. Khoroshilov, Crystallization and glass transition of the diols and aminoalcohols, according to DSC data, *Russ. J. Phys. Chem.* 89 (2015) 910-913. https://doi.org/10.1134/S0036024415050301.

[5] CRC Handbook of Chemistry and Physics, 85th Edition, D.R. Lide, Ed., 2004.

[6] M.N. Islam; M.M. Islam; M.N. Yeasmin, Viscosity of aqueous solutions of 2-methoxyethanol, 2-ethoxyethanol, and ethanolamine, *J. Chem. Thermodyn.* 36 (2004) 889−893. https://doi.org/10.1016/j.jct.2004.06.004.

[7] F. Kermanpour; H.Z. Niakan, Experimental excess molar properties of binary mixtures of (3-amino-1-propanol+isobutanol, 2-propanol) at *T*=(293.15 to 333.15) K and modelling the excess molar volume by Prigogine–Flory–Patterson theory, *J. Chem. Thermodyn.* 54 (2012) 10−19. https://doi.org/10.1016/j.jct.2012.02.036.

[8] F. Kermanpour, Z.G. Kheyrabadi, Experimental Study of Some Thermodynamic Properties of Binary Mixtures Containing 3-Amino-1-propanol, 2-Aminoethanol, and 1-Butanol at Temperatures of 293.15–333.15 K to Model the Excess Molar Volumes Using the PFP Theory, *J. Chem. Eng. Data* 65 (2020) 5360-5368. https://doi.org/10.1021/acs.jced.0c00512.

[9] J. Bentes, A. García-Abuín, A.G. Gomes, D. Gómez-Díaz, J. M. Navaza, A. Rumbo, $CO_2$ chemical absorption in 3-amino-1-propanol aqueous solutions in BC reactor, *Fuel Proc. Technology* 137 (2015) 179-185. https://doi.org/10.1016/j.fuproc.2015.03.030.

[10] L. Dong, J. Chen, G. Gao, Solubility of Carbon Dioxide in Aqueous Solutions of 3-Amino-1-propanol, *J. Chem. Eng. Data* 55 (2010) 1030-1034. https://doi.org/10.1021/je900492a.

[11] A. Henni, J. Li, P. Tontiwachwuthikul, Reaction Kinetics of $CO_2$ in Aqueous 1-Amino-2-Propanol, 3-Amino-1-Propanol, and Dimethylmonoethanolamine Solutions in the Temperature Range of 298−313 K Using the Stopped-Flow Technique, *Ind. Eng. Chem. Res.* 47 (2008) 2213-2220. https://doi.org/10.1021/ie070587r.

[12] A.V. Rayer, K.Z. Sumon, A. Henni, P. Tontiwachwuthikul, Kinetics of the reaction of carbon dioxide ($CO_2$) with cyclic amines using the stopped-flow technique, *Energy Procedia* 4 (2011) 140-147. https://doi.org/10.1016/j.egypro.2011.01.034.

[13] W. Liu, W.Liu, S. Dai, T. Yang. Z. Li, P. Fang, Enhancing the purity of magnesite ore powder using an ethanolamine-based collector: Insights from experiment and theory, *J. Mol. Liq.* 268 (2018) 215-222. https://doi.org/10.1016/j.molliq.2018.07.067.

[14] Yu.V. Novakovskaya, Conjugation in hydrogen-bonded systems, *Struct. Chem.* 23 (2012) 1253-1266. https://doi.org/10.1007/s11224-012-0029-8.

[15] D.L. Thomsen, J.L. Axson, S.D. Schroder, J.R. Lane, V. Vaida, H.G. Kjaergaard, Intramolecular Interactions in 2-Aminoethanol and 3-Aminopropanol, *J. Phys. Chem. A* 117 (2013) 10260-10273. https://doi.org/10.1021/jp405512y.

[16] A.S. Khalil, R.J. Lavrich, Intramolecular hydrogen bond stabilized conformation of 3-aminopropanol, *J. Mol. Spectrosc.* 370 (2020) 111279. https://doi.org/10.1016/j.jms.2020.111279.

[17] A.A. Granovsky, *Firefly version 8.2*, www http://classic.chem.msu.su/gran/firefly/index.html.

[18] *Chemcraft - graphical software for visualization of quantum chemistry computations*, https://www.chemcraftprog.com.

[19] Yu.V. Novakovskaya, M.N. Rodnikova, Monoethanolamine Clusters as Prototypes of the Inherent Structure Elements of the Liquid, *Doklady Phys. Chem.* 467 (2016), 60-62. https://doi.org/10.1134/S0012501616040047.

[20] N.K. Balabaev, Yu.V. Novakovskaya, M.N. Rodnikova, Repeated Elements of the Structure of Liquid Monoethanolamine, *Doklady Phys. Chem.* 479 (2018) 47-51. https://doi.org/10.1134/S0012501618030016.